\documentclass[aps,prd,reprint,twocolumn,superscriptaddress,showpacs]{revtex4-1}

\usepackage{graphicx}
\usepackage{mathrsfs}
\usepackage{bm}
\usepackage{amsmath}
\usepackage{dcolumn}
\usepackage{epstopdf}
\usepackage{dsfont}
\usepackage{amssymb}
\usepackage{tabularx}
\usepackage{array}
\usepackage{float}
\usepackage{color}
\usepackage{epstopdf}
\usepackage{mathrsfs}
\usepackage[colorlinks, linkcolor=blue,anchorcolor=blue,citecolor=blue,urlcolor=blue]{hyperref}

\begin{document}
	
\title{Correlation-driven topological and valley states in monolayer VSi$_{2}$P$_{4}$}

\author{Si Li}\email{sili@nwu.edu.cn}
\affiliation{School of Physics, Northwest University, Xi'an 710069, China}

\author{Qianqian Wang}
\affiliation{Research Laboratory for Quantum Materials, Singapore University of Technology and Design, Singapore 487372, Singapore}

\author{Chunmei Zhang}
\affiliation{School of Physics, Northwest University, Xi'an 710069, China}

\author{Ping Guo}
\affiliation{School of Physics, Northwest University, Xi'an 710069, China}

\author{Shengyuan A. Yang}
\affiliation{Research Laboratory for Quantum Materials, Singapore University of Technology and Design, Singapore 487372, Singapore}
	
\begin{abstract}

Electronic correlations could have significant impact on the material properties. They are typically pronounced for localized orbitals and enhanced in low-dimensional systems, so two-dimensional (2D) transition metal compounds could be a good platform to study their effects.
Recently, a new class of 2D transition metal compounds, the MoSi$_2$N$_4$-family materials, have been discovered, and some of them exhibit intrinsic magnetism. Here, taking monolayer VSi$_{2}$P$_{4}$ as an example from the family,
we investigate the impact of correlation effects on its physical properties, based on the first-principles calculations with the DFT$+U$ approach.
We find that different correlation strength can drive the system into a variety of interesting ground states, with rich magnetic, topological and valley features. With increasing correlation strength, while the system favors a ferromagnetic semiconductor state for most cases, the magnetic anisotropy and the band gap type undergo multiple transitions, and in the process, the band edges can form single, two or three valleys for electrons or holes. Remarkably, there is a quantum anomalous Hall (QAH) insulator phase, which has a unit Chern number and has its chiral edge states polarized in one of the valleys. The boundary of the QAH phase correspond to the half-valley semimetal state with fully valley polarized bulk carriers. We further show that for phases with the out-of-plane magnetic anisotropy, the interplay between spin-orbit coupling and orbital character of valleys enable  an intrinsic valley polarization for electrons but not for holes. This electron valley polarization can be switched by reversing the magnetization direction, providing a new route of magnetic control of valleytronics. Our result sheds light on the possible role of correlation effects in the 2D transition metal compounds, and it will open new perspectives for spintronic, valleytronic and topological nanoelectronic applications based on these materials.

\end{abstract}
	
\maketitle
\section{Introduction}

The impact of electronic correlations on material properties, especially on the magnetic, topological, and valley properties, has been a fascinating subject of research~\cite{xiang2007cooperative,wan2011topological,wang2015interaction,gray2016correlation,sorella2018correlation,leonov2015correlation,chen2019interaction,cui2020correlation,choi2021correlation}. The correlation effects are typically strong in transition metal elements with localized $d$ electrons, and the effects would be further enhanced with reduced dimensionality. Therefore, two-dimensional (2D) transition metal compounds could offer good opportunities to explore the manifestations of electronic correlation effects.

Recently, a new class of 2D transition metal compounds, the MoSi$_2$N$_4$-family materials, have been discovered~\cite{hong2020chemical}. Some members such as monolayer
MoSi$_2$N$_4$ and WSi$_2$N$_4$ were successfully synthesized in experiment~\cite{hong2020chemical}, and more than 60 ternary compounds with similar structures were predicted to be stable~\cite{wang2021intercalated}. A variety of interesting physical properties were suggested for this family of materials, including the Dirac valley structures and valley-contrasting properties, intrinsic magnetism, and nontrivial band topology~\cite{hong2020chemical,li2020valley,wang2020unexpected,wang2021intercalated,yang2021valley,ai2021theoretical,zhong2021strain,mortazavi2021exceptional,guo2020coexistence,wu2021semiconductor,cui2021spin,akanda2021magnetic,islam2021tunable}. On the other hand, the possible impact of correlation effects on these properties, which could be significant in this material family, has not been clearly understood yet.

In this work, we investigate this problem and reveal the importance of electronic correlation in determining the magnetic, topological, and valley properties.
We take monolayer VSi$_2$P$_4$ as an example and perform a detailed study based on the first-principles calculations on the DFT+$U$ level.
We choose VSi$_2$P$_4$ because first, the 3$d$ element V typically has strong correlations, as manifested in the famous examples of VO$_2$, V$_2$O$_3$, and vanadium oxides with Magn\'{e}li phases~\cite{khomskii2014transition}; second, monolayer VSi$_2$P$_4$ has magnetism and interesting valley structures, whereas the previous studied MoSi$_2$N$_4$, WSi$_2$N$_4$, and MoSi$_2$As$_4$ are non-magnetic~\cite{wang2021intercalated,li2020valley}.
We find that different Hubbard $U$ strengths can drive the system into different ground states, enabling a rich phase diagram (see Fig.~\ref{fig4}). We show that the material is an indirect gap ferromagnetic semiconductor at small $U$. With increasing $U$ strength, it first changes to direct gap with a pair of Dirac type valleys at $K$ and $K'$ points of the Brillouin zone (BZ). Then the magnetic anisotropy switches from in-plane to out-of-plane at $U\sim 2.13$ eV. Remarkably, between $U\sim 2.25$ and 2.36 eV, the system transitions into a quantum anomalous Hall (QAH) insulator phase featured by a unit Chern number and chiral edge states polarized in a single valley. The boundaries of this QAH phase correspond to the critical semimetal states where the gap only closes at one of the two valleys. Further increasing $U$ will drive  two additional transitions in the magnetic anisotropy and a transition from direct to indirect gap. Importantly, we show that the out-of-plane ferromagnetic semiconductor phases have an intrinsic valley polarization for electrons but not for holes. We explain this asymmetry from the different orbital contributions in the spin-orbit coupling (SOC). In such a state, the electron valley polarization can be switched by reversing the magnetization direction, providing a new route of magnetic control of valleytronics. Experimental signatures and possible ways to tune the correlation strength are discussed. Our result highlights the role of correlation effects in the 2D MoSi$_2$N$_4$-family materials and deepens our understanding of interesting correlation-driven topological and valley states.

\section{Computation METHODS}

We performed first-principles calculations based on the density functional theory (DFT), using the projector augmented wave method as implemented in the Vienna \emph{ab initio} simulation package~\cite{Kresse1994,Kresse1996,PAW}. The generalized gradient
approximation (GGA) with the Perdew-Burke-Ernzerhof (PBE)~\cite{PBE} realization was adopted for the exchange-correlation functional. The cutoff energy was chosen as 500 eV, and the BZ was sampled with a $\Gamma$-centered $k$ mesh of size $15\times 15\times 1$. The energy and force convergence criteria were set to be $10^{-6}$ eV and $0.001$ eV/\AA, respectively. A vacuum layer with a thickness of 20 \AA\ was taken to avoid artificial interactions between periodic images. The phonon
spectrum was calculated using the PHONOPY code through the
density functional perturbation theory (DFPT) approach~\cite{togo2015first}, with a $2\times 2\times 1$ supercell
and a $8\times 8\times 1$ $q$-grid (the total energy converges
with an accuracy of $10^{-7}$ eV). The correlation effects for the V-3$d$ electrons were treated by the DFT$+U$ method~\cite{anisimov1991,dudarev1998}. The band structure was also calculated by using the Heyd-Scuseria-Ernzerhof hybrid functional method (HSE06)~\cite{heyd2003hybrid} (see the Supplemental Material~\cite{SM}). The Berry curvature and the intrinsic anomalous Hall conductivity were evaluated using the WANNIER90 package~\cite{mostofi2008wannier90,wang2006ab}. The edge states were calculated by using the iterative Green function method~\cite{Green}, as implemented in the WannierTools package~\cite{wu2018wanniertools}. Curie temperatures were estimated by using the Monte Carlo simulations as implemented in the VAMPIRE atomistic simulation package~\cite{evans2014atomistic}.

\section{Structure and magnetism}

\begin{figure}[htbp]
	\includegraphics[width=8.5cm]{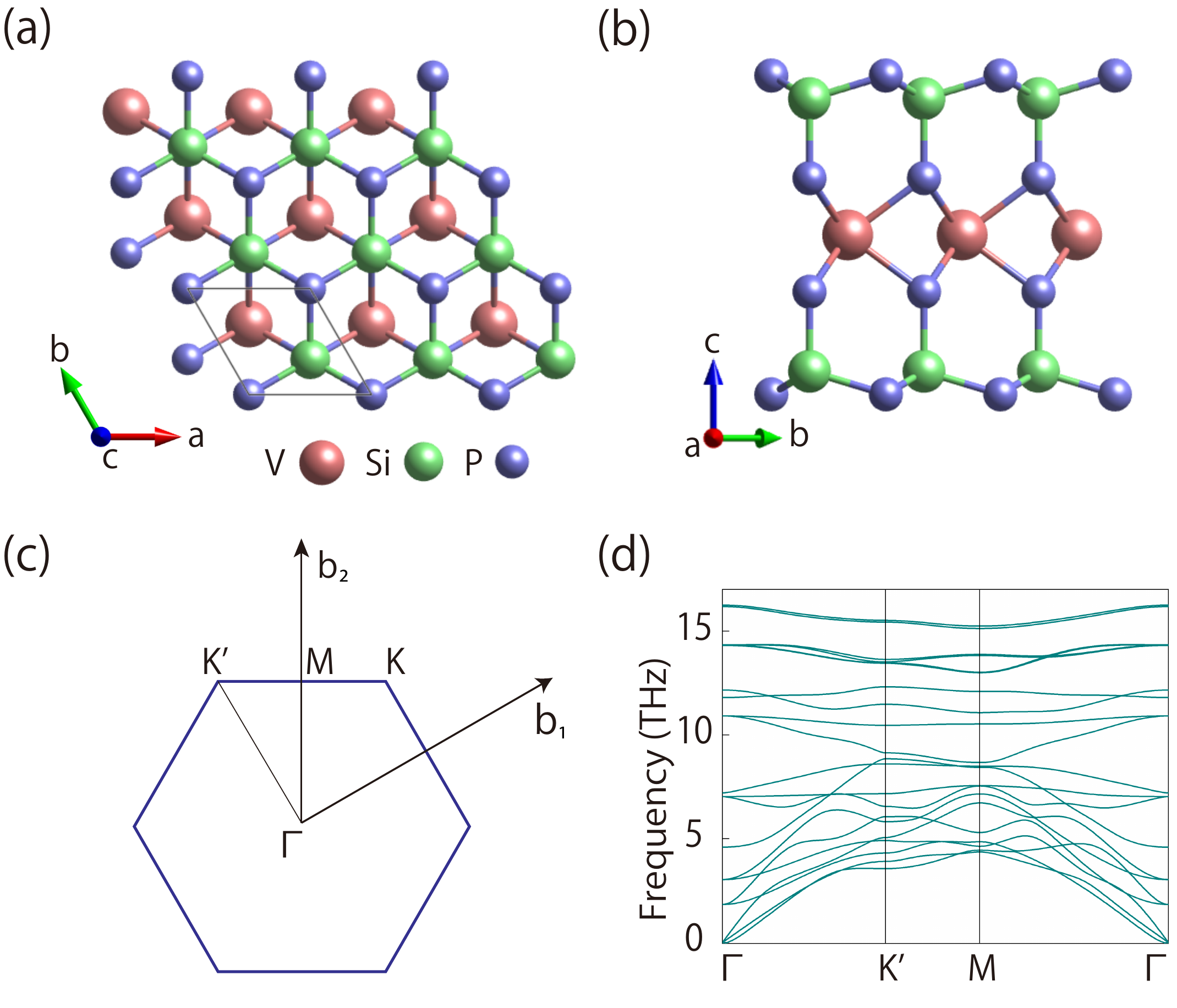}
	\caption{(a) Top and (b) side view of the crystal structure of monolayer VSi$_{2}$P$_{4}$. The primitive cell is marked by the solid lines in (a). (c) Brillouin zone (BZ) with high symmetry points labeled. (d) Calculated phonon spectrum for monolayer VSi$_{2}$P$_{4}$.
		\label{fig1}}
\end{figure}

\begin{figure}[htbp]
	\includegraphics[width=6.8cm]{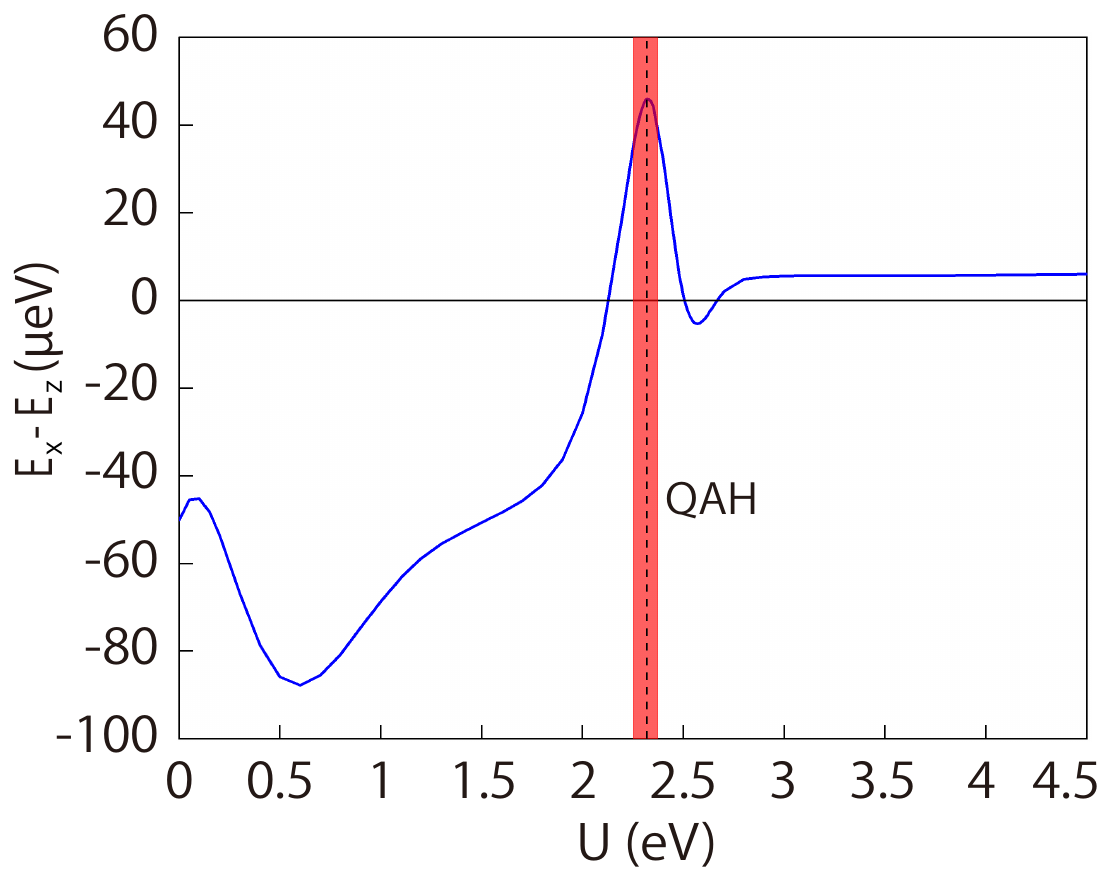}
	\caption{Energy difference between in-plane and out-of-plane magnetization directions for the FM state. The red colored region indicates the QAH phase.
		\label{fig2}}
\end{figure}

\begin{figure}[b]
	\includegraphics[width=6.8cm]{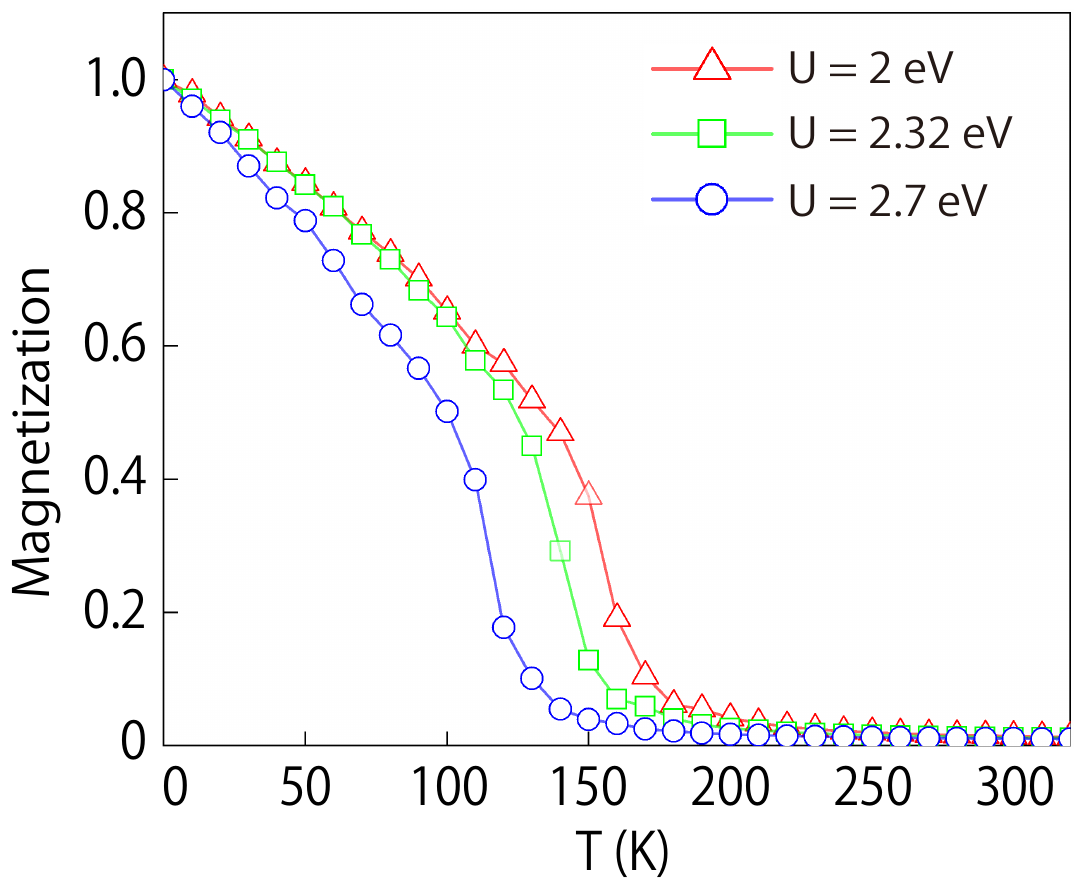}
	\caption{Normalized magnetic moment as a function of temperature by Monte Carlo simulations with different $U$ values.
		\label{fig3}}
\end{figure}

We construct the monolayer VSi$_{2}$P$_{4}$ lattice model with the same structure as the already synthesized monolayer MoSi$_2$N$_4$~\cite{hong2020chemical}. It has the hexagonal lattice structure with space group $P \overline{6} m 2$ (No.~$187$). As shown in Fig.~\ref{fig1}(a) and \ref{fig1}(b), the structure is built up by septuple atomic layers in the sequence of P-Si-P-V-P-Si-P. It is important to note that the structure breaks the inversion symmetry $\mathcal{P}$. From our first-principles calculations, the fully optimized lattice parameters are given by $a =b= 3.486$ \AA. To confirm the stability of the monolayer structure, we perform the phonon spectrum calculation. The obtained spectrum is plotted in Fig.~\ref{fig1}(d),  which shows that there is no soft phonon mode throughout the BZ, indicating that the structure is dynamically stable.

We then investigate the magnetic properties of monolayer VSi$_{2}$P$_{4}$. We compared energies of the nonmagnetic state, the ferromagnetic (FM) state and several antiferromagnetic (AFM) configurations at different $U$ values (see the Supplemental Material for details~\cite{SM}). For V compounds, the typical $U$ values are around 3 to 4 eV, so here we investigate the $U$ value range from 0 to 4.5 eV. 
We find that the monolayer VSi$_{2}$P$_{4}$ always prefers the FM ground state (even up to $U=5$ eV). The magnetic moment on the V site is about 1.1 $\mu_{B}$.  Nevertheless, the magnetic anisotropy changes with the $U$ values. In Fig.~\ref{fig2}, we plot the energy difference $(E_x-E_z)$ as a function of $U$, where $E_{x/z}$ is the energy per unit cell when the magnetization is along the $x/z$ direction. One can observe several transitions in the magnetic anisotropy.
For $U<2.13$ eV and $U$ between 2.51 and 2.67 eV, the system prefers an in-plane FM state; whereas for $U$ between 2.13 and 2.51 eV and $U>2.67$ eV, it favors an out-of-plane FM state. The different magnetic orientations will affect the symmetry of the system, which will in turn have important influence on the electronic properties, as we shall see in a while.

%To determine the magnetic anisotropy energy (MAE) of monolayer VSi$_{2}$P$_{4}$, we calculate the energy difference between different magnetization directions of V atoms when the SOC effect is included. We find that there almost has no difference in energy when magnetization directions along the in-plane. We can define the MAE as the energy difference between magnetization along $x$ direction and magnetization along $z$ direction, and the MAE with the value of $U$ is shown in Fig.~\ref{fig2}(b). One observes that the magnetization prefers an in-plane direction when $U \textless 2.13$ eV and $2.51$ eV $\textless U \textless 2.67$ eV, and it prefers an out-of-plane direction  when $2.13$ eV $\textless U \textless 2.51$ eV and $U \textgreater 2.67$ eV.

We have also estimated the Curie temperature ($T_C$) for the system at three representative $U$ values ($U=2$ eV, 2.32 eV and 2.7 eV). The calculation is performed by using the Monte Carlo simulations based on an effective classical spin model~\cite{evans2014atomistic}:
\begin{equation}\label{Heisenberg}
	H=-\sum_{ i, j} J_{i j} \bm{S}_{i} \cdot \bm{S}_{j}-K\sum_{i}\left(S_{i}^{z}\right)^{2},
\end{equation}
where $\bm S_i$ is the normalized spin vector on the V site $i$, $J_{ij}$ is the exchange coupling constant between sites $i$ and $j$, and $K$ is the site anisotropy strength. For a rough estimation, we include only the nearest-neighbor coupling $J$ in the model and the leading order anisotropy term, where positive (negative) $K$ corresponds to an easy-axis (easy-plane) anisotropy. The details for extracting these model parameters are presented in the Supplemental Material~\cite{SM}. The obtained model parameters are found to be $J=14.1 $ meV and $K= -25.7 $ $\mu$eV for $U=2$ eV, $J=12.6 $ meV and $K= 46.0 $ $\mu$eV for $U=2.32$ eV, and $J=10.5$ meV and $K= 2.03$ $\mu$eV for $U=2.7$ eV. The simulated magnetization versus temperature curves are shown in Fig.~\ref{fig3}. The estimated $T_C$ is about 161 K for $U=2$ eV, 146 K for $U=2.32$ eV, and 120 K for $U=2.7$ eV.
%The relatively high Curie temperature suggests that monolayer VSi$_{2}$P$_{4}$
%can be suitable for practical spintronics applications.

\section{Evolution of electronic structures}

\begin{figure}[b]
	\includegraphics[width=8.8cm]{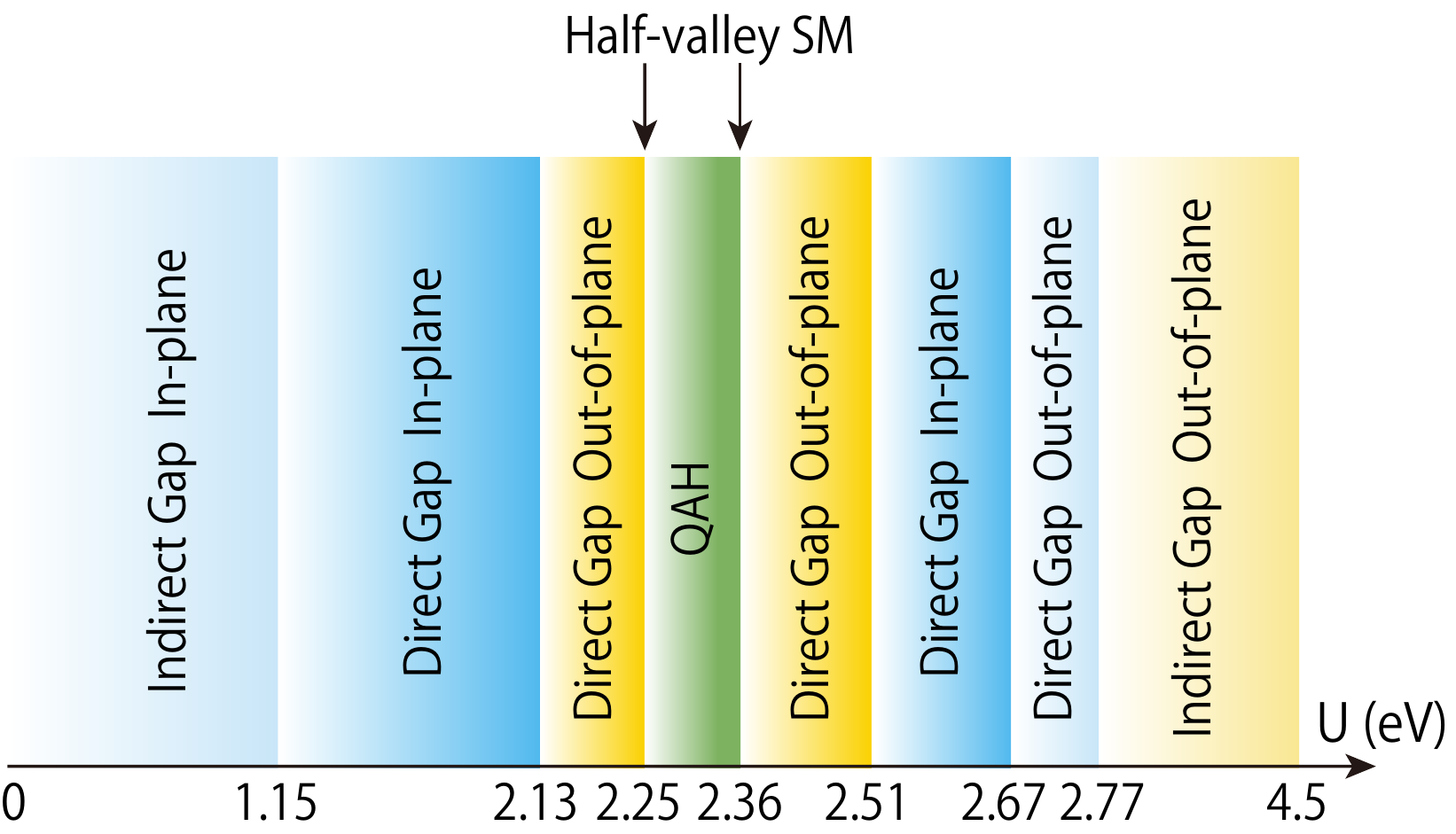}
	\caption{ Phase diagram for monolayer VSi$_{2}$P$_{4}$ with different $U$ values.
		\label{fig4}}
\end{figure}

\begin{figure*}[htbp]
	\includegraphics[width=17.6cm]{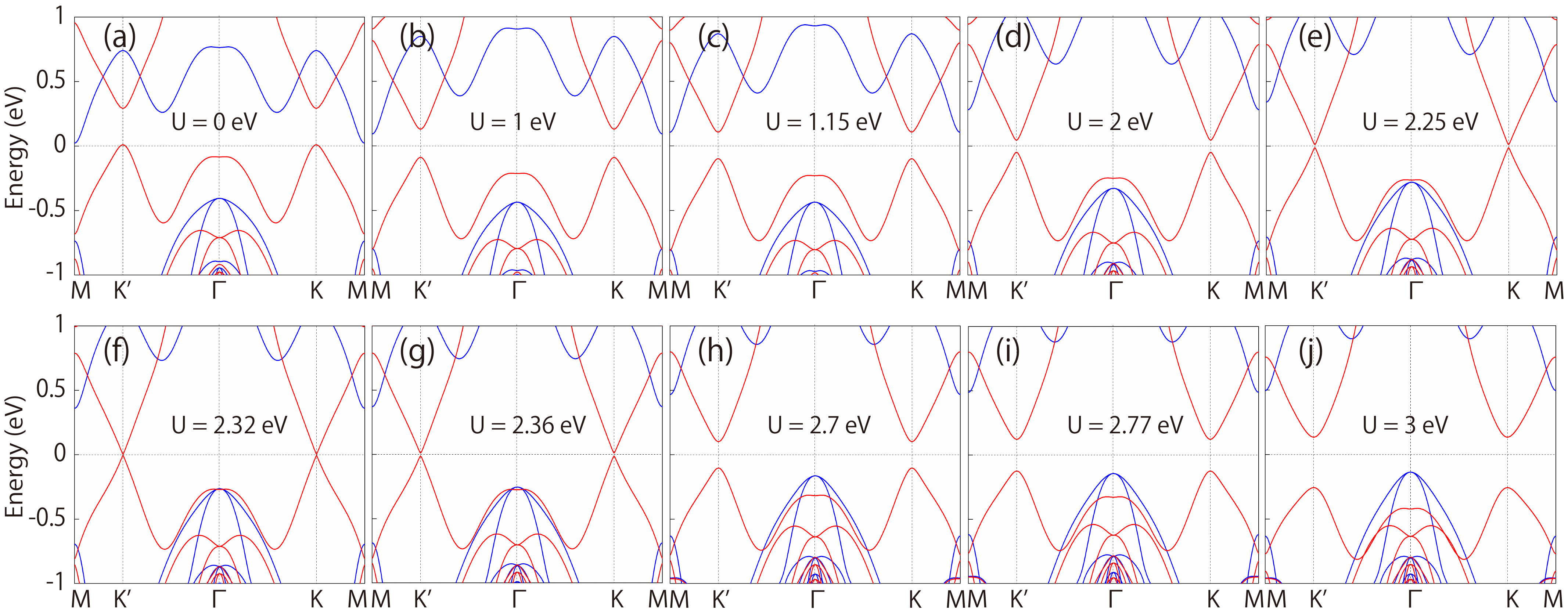}
	\caption{Spin-polarized band structures of monolayer VSi$_{2}$P$_{4}$ obtained from GGA$+U$ ($U$ varies from 0 to 3 eV) without SOC. The red (blue) color represents spin-up (spin-down) bands.
		\label{fig5}}
\end{figure*}

\begin{figure*}[htbp]
	\includegraphics[width=17.6cm]{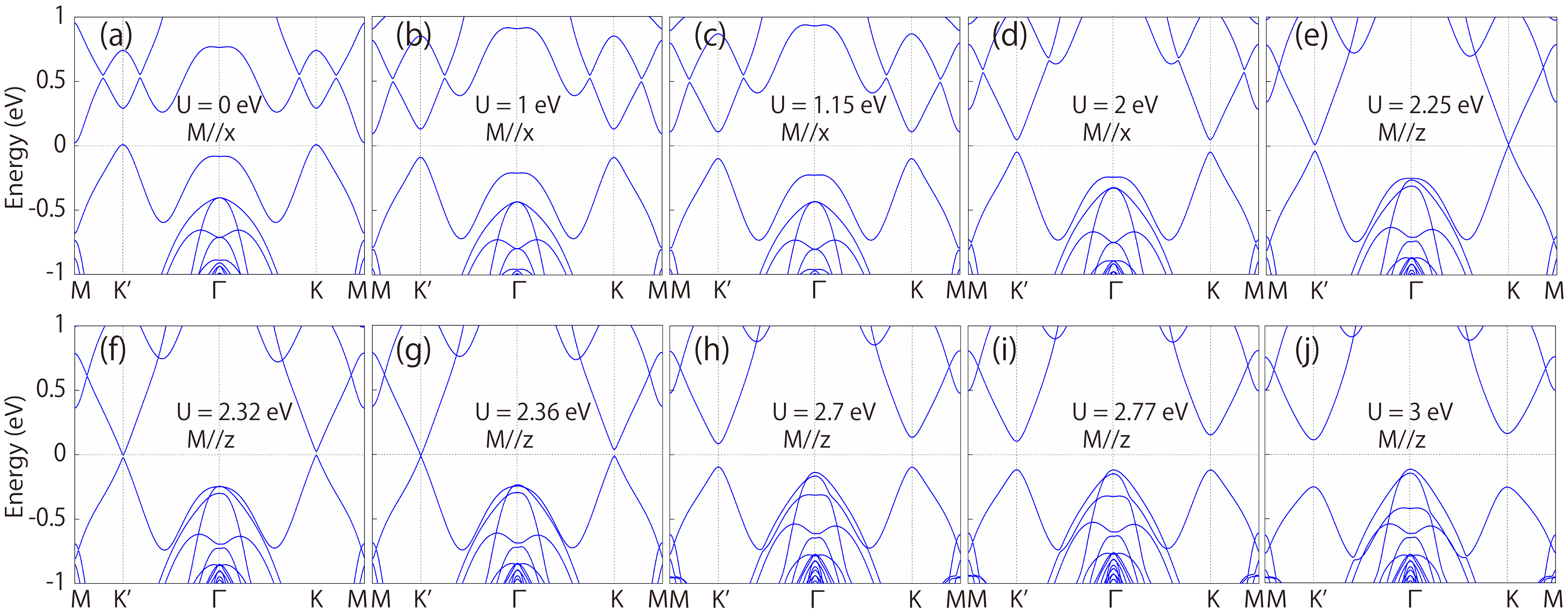}
	\caption{Band structures of monolayer VSi$_{2}$P$_{4}$ obtained from GGA$+U$ ($U$ varies from 0 to 3 eV) with SOC included.
		\label{fig6}}
\end{figure*}

We have investigated the evolution of electronic band structures with $U$ and obtained the phase diagram in Fig.~\ref{fig4}. The representative band structures at different $U$ values without and with SOC are plotted in Fig.~\ref{fig5} and Fig.~\ref{fig6}, respectively.

One observes that at small $U$, the system is an indirect gap semiconductor. The valence band maximum (VBM) is at $K$ and $K'$ points, whereas the conduction band minimum (CBM) occurs at the $M$ points. Interestingly, this state has different valley structures for its electrons and holes. In the absence of SOC, the valence band has two degenerate valleys at $K$ and $K'$, but the conduction band has three degenerate valleys at the $M$ points. Moreover, these valleys are in different spin channels: in Fig.~\ref{fig5}(a), the electron valleys are spin down, whereas the hole valleys are spin up. Including the SOC will break the energy degeneracy of these valleys and also mix the two spin channels.

With increasing $U$, the conduction band at $K$ and $K'$ moves down relative to the original CBM at $M$ [see Fig.~\ref{fig5}(a)]. For $U>1.15$ eV, the system become a direct gap semiconductor with the band gap at $K$ and $K'$ points. The CBM and VBM here form a pair of Dirac type valleys, similar to MoS$_2$ type materials~\cite{huang2013metal,tan2015two,lv2015transition,xiao2012coupled,yao2008valley}, but the distinct point is that these valleys belong to the same spin channel, in contrast to the previously studied cases in monolayer WSi$_2$N$_4$~\cite{li2020valley}.

When $U$ is above 2.13 eV, the magnetic anisotropy changes to out-of-plane. Since the magnetization is a pseudovector, the out-of-plane 
FM preserves the horizontal mirror symmetry $M_z$. 
Because of this preserved $M_z$, each band eigenstate under SOC has a well defined spin eigenvalue, either spin up or spin down. More importantly, the out-of-plane FM breaks all possible vertical mirrors of the system (i.e., mirrors perpendicular to the $x$-$y$ plane), hence allowing a non-vanishing Chern number of the 2D system~\cite{liu2013plane}. From Fig.~\ref{fig6}, one can see that the band gap decreases with $U$. Around $U=2.25$, the gap closes and re-opens at $K$ (but not $K'$). And around $U=2.36$, the similar process happens at the $K'$ valley. This gap closing and re-opening scenario suggests a topological phase transition. Indeed, we verify that the phase for $U=2.32$ eV represents a QAH insulator phase, characterized by a unit Chern number.

\begin{figure}[htbp]
	\includegraphics[width=8.5cm]{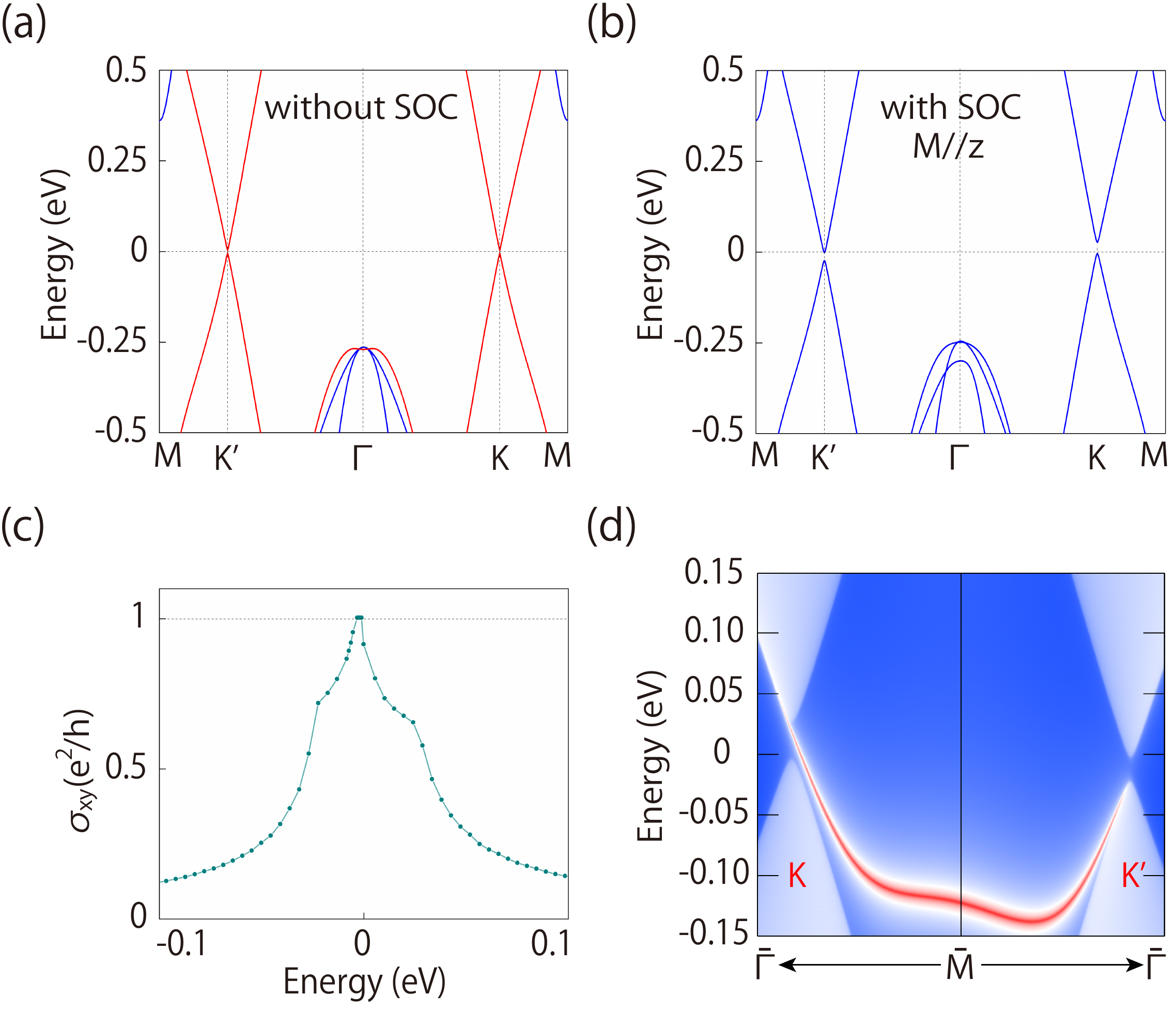}
	\caption{Band structure of monolayer VSi$_{2}$P$_{4}$ with $U=2.32$ eV (a) without SOC (the red and blue colors indicate spin-up and spin-down channels, respectively) and (b) with SOC. (c) Anomalous Hall conductivity versus chemical potential for the case in (b). (d) shows the corresponding edge spectrum for the QAH state in (b). Note that the plot is centered at the $M$ point, so the left valley is $K$ and the right valley is $K'$.
		\label{fig7}}
\end{figure}

We have evaluated the intrinsic anomalous Hall conductivity $\sigma_{xy}^i$ via the first-principles calculations~\cite{jungwirth2002anomalous,yao2004first}. This quantity is given by
\begin{equation}
	\sigma_{x y}^i=-\frac{e^{2}}{h}\frac{1}{2\pi} \int_\text{BZ} {d^2 k}\ \Omega_z\left(\bm k\right).
\end{equation}
$\Omega_z\left(\bm k\right)$ is the Berry curvature of the 2D system.
\begin{equation}
	\Omega_{z}\left(\bm k\right)=-2 \operatorname{Im} \sum_{n\neq n^{\prime}}f_{n\bm k} \frac{\left\langle n \bm{k}\left|v_{x}\right| n' \bm{k}\right\rangle\left\langle n' \bm{k}\left|v_{y}\right| n \bm{k}\right\rangle}{(\omega_{n^{\prime}}-\omega_{n})^{2}},
\end{equation}
where the summation is over both band indices $n$ and $n'$ with $n$ restricted to all occupied bands, $\varepsilon_{n}=\hbar \omega_{n}$ is the band energy, $v$'s are the velocity operators, and $f_{n\bm k}$ is the equilibrium distribution function.
The calculation result for $U=2.32$ eV is shown in Fig.~\ref{fig7}(c). One observes that the anomalous Hall conductivity is $e^2/h$ in the gap, confirming that it is a QAH insulator with a Chern number $\mathcal{C} =1$. One feature of the QAH insulator is the existence of chiral
edge states. In Fig.~\ref{fig7}(d), we plot the corresponding edge spectrum. One observes that there is a single gapless chiral edge band crossing the band gap at the $K$ valley, which is consistent with the quantized QAH conductivity.

It should be noted that the QAH phase here coexists with a valley structure~\cite{pan2014valley}. This leads to several interesting features. First, the gapless chiral edge band acquires a valley character. For example, the edge states in Fig.~\ref{fig7}(d) belong to the $K$ valley. Such chiral edge states could be useful for valleytronics applications, as demonstrated in Ref.~\cite{pan2015valley}. Second, the boundary of the QAH phase at $U=2.25$ and 2.36 eV are critical points of topological phase transitions. The two critical states must be gapless. As mentioned, these two states have band gap closed only at one of the two valleys. In this case, the transport in the bulk would also be fully valley polarized. Such interesting states correspond to the concept of half-valley metal recently proposed in Ref.~\cite{hu2020concepts}.

Further increasing $U$ will drive another two transitions in the magnetic anisotropy. For $U$ between 2.51 and 2.67 eV, the magnetization becomes in-plane; and for $U>2.67$ eV, the magnetization switches to out-of-plane. In addition, at $U\sim 2.77$ eV, the band gap changes from direct to indirect types. This is accompanied with the switch of VBM from $K$ and $K'$ to $\Gamma$. As shown in Fig.~\ref{fig8}, the valence band at $\Gamma$ belong to the spin down channel and is dominated by the $d_{xz}$ and $d_{yz}$ orbital, which is distinct from the states at $K$ and $K'$.

\section{Magnetic valley control}

\begin{figure}[htbp]
	\includegraphics[width=8.5cm]{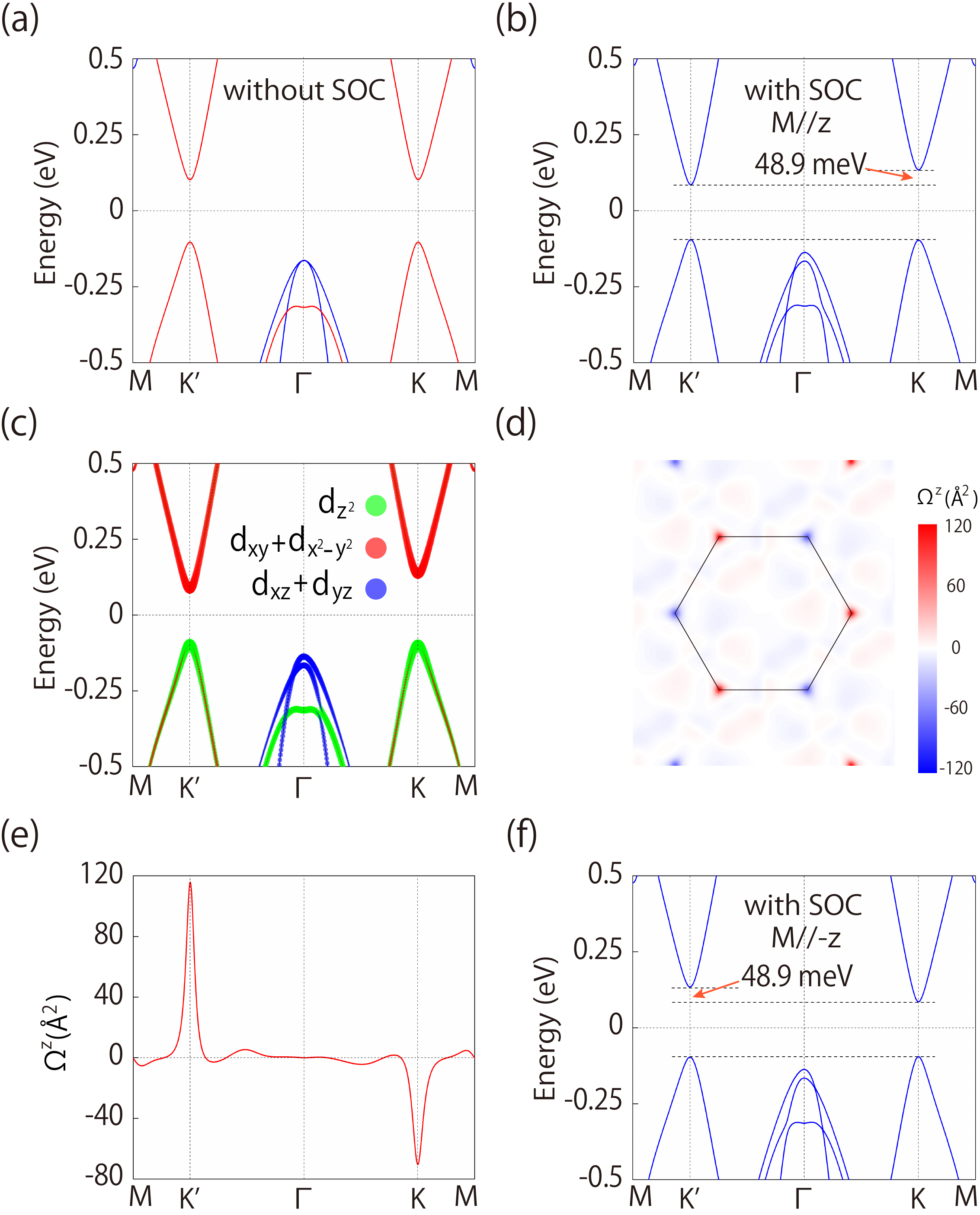}
	\caption{ Band structure of monolayer VSi$_{2}$P$_{4}$ with $U=2.7$ eV (a) without SOC and (b) with SOC. Here, the magnetization is along $+z$ direction. (c) Orbital-projected band structure for the case in (b). (d) and (e) show the distribution of Berry curvature for the valence bands. (f) Band structure of monolayer VSi$_{2}$P$_{4}$ when the magnetization is switched to the $-z$ direction.
		\label{fig8}}
\end{figure}

We note that in the phase diagram Fig.~\ref{fig4}, there are several regions with out-of-plane FM and with the low-energy physics occurring at the $K$ and $K'$ valleys. A representative state is shown in Fig.~\ref{fig8} with $U=2.7$ eV. Here, we zoom in the two valleys and compare the band structures without and with SOC. Without SOC, the valleys are degenerate in energy for both conduction and valence bands. Interestingly, after including SOC, the valley degeneracy for the conduction band is broken, the conduction valley at $K$ is about 48.9 meV higher than that at $K'$. Meanwhile, the degeneracy for the valence band valleys is almost unaffected [see Fig.~\ref{fig8}(b)].

To understand this peculiar effect of SOC on the band structure, we first note that the VBM and CBM in Fig.~\ref{fig8}(c) are dominated by different orbital components. The VBM is mainly from the V $d_{z^2}$ orbitals, where the CBM is dominated by the V $d_{xy}$ and $d_{x^2-y^2}$ orbitals. At $K$ and $K'$, the little group is $C_{3h}$ for the out-of-plane magnetization. Hence, the
orbital basis for VBM and CBM at the two valleys can be chosen as  $\left|\psi_{v}^{\tau}\right\rangle=\left|d_{z^{2}}\right\rangle\otimes |\uparrow\,\rangle$ and $\left|\psi_{c}^{\tau}\right\rangle=\frac{1}{\sqrt{2}}\left(\left|d_{x^{2}-y^{2}}\right\rangle+i \tau\left|d_{x y}\right\rangle\right)\otimes|\uparrow\,\rangle$, where $\tau=\pm 1$ is the valley index corresponding to $K$/$K'$ and from Fig.~\ref{fig8}(a), we know that VBM and CBM belong to the same spin channel.

The effect of SOC on the VBM and CBM states may be approximated by the perturbation term
\begin{equation}
  \hat{H}_\text{SOC}=\lambda \hat{S} \cdot \hat{L},
\end{equation}
where $\lambda$ is the coupling strength, $\hat{S}$ and $\hat{L}$ are the spin and orbital angular momentum operator, respectively. This SOC term can be formally expressed as  $\hat{H}_\text{SOC}=\hat{H}_\text{SOC}^{0}+\hat{H}_\text{SOC}^{1}$, with~\cite{whangbo2015prediction,whangbo2019electronic,peng2020intrinsic}
\begin{equation}
\begin{aligned}
\hat{H}_\text{SOC}^{0}=& \lambda \hat{S}_{z^{\prime}}\left(\hat{L}_{z} \cos \theta+\frac{1}{2} \hat{L}_{+} e^{-i \phi} \sin \theta+\frac{1}{2} \hat{L}_{-} e^{+i \phi} \sin \theta\right),\\
\hat{H}_\text{SOC}^{1}=& \frac{\lambda}{2}\left(\hat{S}_{+^{\prime}}+\hat{S}_{-^{\prime}}\right) \\ & \times\left(-\hat{L}_{z} \sin \theta+\frac{1}{2} \hat{L}_{+} e^{-i \phi} \cos \theta+\frac{1}{2} \hat{L}_{-} e^{+i \phi} \cos \theta\right),
\end{aligned}
\end{equation}
where in the most general case, $\hat{L}$ and $\hat{S}$ may be expressed in different coordinate systems $(x, y, z)$ and $\left(x^{\prime}, y^{\prime}, z^{\prime}\right)$, respectively, $\theta$ and $\phi$ are the polar angles that relate the two systems, $\hat{L}_{\pm}=\hat{L}_{x}\pm i\hat{L}_{y}$, and $\hat{S}_{{\pm}^{\prime}}=\hat{S}_{{x}^{\prime}}\pm i\hat{S}_{{y}^{\prime}}$. For out-of-plane magnetization, it is
convenient to take the two coordinate systems identical, then we have $\theta=\phi=0$. Since the VBM and CBM here are in the same spin channel, the first order perturbation from $\hat{H}_\text{SOC}^{1}$ vanishes identically. It follows that to first order in $\lambda$, the SOC term can be reduced to
\begin{equation}
  \hat{H}_\text{SOC}=\lambda \hat{S}_z \hat{L}_z.
\end{equation}
The resulting energy shifts for the VBM and CBM at the two valleys are given by
$E_{v}^{\tau}=\langle\psi_{v}^{\tau}|\hat{H}_{\text {SOC }}| \psi_{v}^{\tau}\rangle$ and $E_{c}^{\tau}= \langle\psi_{c}^{\tau}|\hat{H}_{\mathrm{soc}}| \psi_{c}^{\tau}\rangle$, respectively. Consequently, the energy difference between valleys $K$ and $K'$ is given by
\begin{equation}
\begin{aligned}
	E_{c}^{K}-E_{c}^{K'}= & i\left\langle d_{x^{2}-y^{2}}\left|\hat{H}_{\mathrm{SOC}}\right| d_{x y}\right\rangle\\&-i\left\langle d_{x y}\left| \hat{H}_{\mathrm{SOC}}\right|d_{x^{2}-y^{2}}\right\rangle
	  \approx  4 \lambda,\\
	E_{v}^{K}-E_{v}^{K'}= & 0,
\end{aligned}
\end{equation}
where we have used that $\hat{L}_{z}\left|d_{x^{2}-y^{2}}\right\rangle=2 i\left|d_{x y}\right\rangle$ and $\hat{L}_{z}\left|d_{x y}\right\rangle=-2 i\left|d_{x^{2}-y^{2}}\right\rangle$~\cite{khomskii2014transition}. This analysis demonstrates that to the first order in SOC, the valley degeneracy splits for the conduction band but not the valence band, consistent with our first-principles results.

Actually, following the similar approach as in Ref.~\cite{xiao2012coupled,li2020valley}, we can obtain the following low-energy effective model for the state in Fig.~\ref{fig8}(b)
\begin{equation}
  H_\text{eff}=\alpha(\tau k_x\sigma_x+k_y\sigma_y)+\frac{\Delta}{2}\sigma_z+\lambda m_z \tau(\sigma_z+1),
\end{equation}
where $k$ is measured from each valley center, $\alpha$ is a model parameter, $\sigma$'s are the Pauli matrices in the two basis at each valley, $\Delta$ is the band gap in the absence of SOC, and $m_z=\pm 1$ denotes the magnetization direction along the $\pm z$ direction.

Importantly, the valley polarization for electrons would be switched by reversing the magnetization direction $m_z$. This is confirmed by calculation result, as shown in Fig.~\ref{fig8}(f).  Moreover, since the low-energy bands at $K$ and $K'$ belong to the same spin channel (the spin majority channel),
the spin polarization of the carriers is simultaneously switched. The Magnetic control of valley polarization in non-magnetic materials is through the coupling between the orbital magnetic moment and the external magnetic field~\cite{cai2013magnetic}. Here, the mechanism is different. It is through the coupling among intrinsic magnetism, valley and SOC. In practice, using the intrinsic magnetism rather than applied magnetic field allows a ``nonvolatile'' scheme for generating valley polarization. In addition, magnetism can be controlled in a fully electric manner, e.g., by using current pulses through spin torques, which is desired for device applications. Thus, the mechanism discussed here offers a new route for controlling the valley and spin degrees of freedom.

\section{Discussion and Conclusion}

We have demonstrated the importance of electron correlations on the physical properties of monolayer VSi$_2$P$_4$. The different $U$ values can result in different ground states, with intriguing magnetic, valley, and topological features. Of course, for a given material, the correlation strength is fixed, and the material should belong to a particular phase in the phase diagram. For example, the self-consistent procedure based on a linear response approach gives an estimation of $U=4$ eV for this material, which would put it into the phase of an indirect gap FM semiconductor with valley polarization in the conduction band~\cite{wang2021intercalated}. Meanwhile, result from the hybrid functional approach (HSE06) shows qualitative features similar to that of $U\sim 2.7$ eV~\cite{SM}. Due to its sensitivity to correlation strength, the actual phase of the material should be determined from future experiment.

Nevertheless, we wish to point out that the physics of the rich phase diagram and the phase transitions can still be exhibited in practice. 
An essential point of correlated systems is that the physics depends on the competition between kinetic and interaction energies. By suppressing/enhancing the kinetic energy (e.g., by applied strain), one then \emph{effectively} enhances/suppresses the correlation effect.
For example, applied strain or pressure can modify the band width, which effectively controls the relative importance of electronic correlations. To demonstrate this point, consider the monolayer VSi$_2$P$_4$ with $U=2$ eV. According to Fig.~\ref{fig4}, the ground state should be trivial FM semiconductor. By applying a 1.8\% biaxial strain on the system, we find that the system is driven into the QAH phase. This conforms with out expectation: The tensile strain tends to suppress the kinetic energy of electrons and make the electrons more localized, which effectively enhances electronic correlation. Following this discussion, there could be versatile phase transitions in this 2D material family driven by strain, pressure, temperature, and etc.

We take monolayer VSi$_2$P$_4$ as a concrete example in this study. It is clear that the analysis here can be readily extended to other members of the MoSi$_2$N$_4$ family. It was noted in Ref.~\cite{wang2021intercalated} that there are several variants in lattice structures for this family. For example, 2D VSi$_2$P$_4$ may also be stabilized in another predicted $\delta_4$ structure~\cite{wang2021intercalated}. Nevertheless, the important point is that this family of materials share the same structural motif, i.e., the central triangular lattice plane of the transition metal elements. The low-energy states as well as the electronic correlations are dominated by this plane. Thus, one can naturally expect that the rich correlation-driven physics should be common among the different modified structures for this family. Yet, the effective correlation strength should vary from material to material. Generally, the $d$-electrons become less correlated when going from 3$d$ to $4d$ and 5$d$ series, because of the increasing covalency with surrounding ions and the increasing band width.

In conclusion, we have demonstrated the rich correlation driven physics in monolayer VSi$_{2}$P$_{4}$, as a representative of the 2D MoSi$_2$N$_4$ material family. We show that different correlation strength characterized by the $U$ values can result in a rich phase diagram, with interesting interplay between magnetic, valley, and topological features. Particularly, we observe multiple transitions in the magnetic anisotropy, valley structure, and band gap type. There exists a QAH phase characterized by a unit Chern number. The boundary of the QAH phase corresponds to the half-valley semimetal state with fully valley polarized carriers. We show that even for trivial semiconductor phase with out-of-plane magnetization, there is a valley polarization generated by SOC. This polarization exists for electrons but not for holes, and it can be switched by reversing the magnetization. This study deepens our understanding of the correlation effects in the 2D MoSi$_2$N$_4$ family materials, and it will open new perspectives for spintronic, valleytronic and topological nanoelectronic applications based on these materials.

\begin{acknowledgements}
The authors thank D. L. Deng for valuable discussions. This work is supported by the National Natural Science Foundation (NSF) of China (Grants No. 12004306 and No. 11974277), the Singapore Ministry of Education AcRF Tier 2 (Grant No.~MOE2019-T2-1-001).
\end{acknowledgements}

\emph{Note added:} Recently another work appeared, which reported similar magnetic and valley physics in a related material, the monolayer VSi$_2$N$_4$~\cite{zhou2021valley}.

%\begin{appendix}
%\end{appendix}

%\bibliography{VSi2P4_ref}

%merlin.mbs apsrev4-1.bst 2010-07-25 4.21a (PWD, AO, DPC) hacked
%Control: key (0)
%Control: author (8) initials jnrlst
%Control: editor formatted (1) identically to author
%Control: production of article title (-1) disabled
%Control: page (0) single
%Control: year (1) truncated
%Control: production of eprint (0) enabled
%

\end{document}